\documentclass[10pt, aps, prd, superscriptaddress, nofootinbib, showpacs, twocolumn]{revtex4-2}
 \pdfoutput=1
\usepackage[english]{babel}
\usepackage{amsmath, amssymb, amsfonts, amsthm, mathrsfs}
\usepackage[all]{xy}
\usepackage{bm}
\usepackage{stmaryrd}
\usepackage{graphicx}
\usepackage[colorlinks]{hyperref}
\usepackage{color}

\newcommand{\calA}{\mathcal A}
\newcommand{\calF}{\mathcal F}
\newcommand{\calL}{\mathcal L}
\newcommand{\scrA}{\mathscr A}
\newcommand{\scrF}{\mathscr F}

\DeclareMathOperator{\tr}{tr}

\begin{document}

\title{New ``metric-affine-like'' generalization of Yang-Mills theory}

\author{W\l adys\l aw Wachowski}
\email{vladvakh@gmail.com}
\affiliation{I.\,E. Tamm Theory Department, P.\,N. Lebedev Physical Institute of the Russian Academy of Sciences, Leninsky Prospekt 53, Moscow 119991, Russia}

\begin{abstract}
We suggest a new generalization of the $\mathrm{U}(n)$ Yang-Mills theory obtained by relaxing the condition of covariant constancy of the Hermitian form in the fibers, $\nabla_a g_{\alpha\beta'} \ne 0$. This theory is a simpler analogue of the metric-affine gravity with $\nabla_a g_{bc} \ne 0$. In our case, connection $\nabla_a$ and Hermitian form $g_{\alpha\beta'}$ are two independent variables so total curvature and total potential are no longer anti-Hermitian matrices: thus, along with the standard YM potential $\bm{A}_a$ and field strength tensor $\bm{F}_{ab}$, it contains non-trivially interacting fields $\bm{B}_a$, $\bm{h}$, and $\bm{G}_{ab}$, $\bm{N}_a$, forming a non-Abelian generalization of St\"{u}ckelberg theory. Due to the spontaneous symmetry breaking $\mathrm{GL}(n,\mathbb{C}) \to \mathrm{U}(n)$, these new fields can be made massive, and the limit $M\to\infty$ restores the standard YM theory.
\end{abstract}

\maketitle
\begingroup
  \hypersetup{linkcolor=blue}
  \tableofcontents
\endgroup

\section{Motivation}

The deep kinship between the Yang-Mills theory (YM), which describes strong and electro-weak interactions, and the Einstein theory of gravity (EG) is well-known today. This kinship is based on the fact that at the center of both theories lies the same geometric object---covariant derivative, or connection. In case of YM, it has the form
\begin{equation} \label{CovDerYM}
\nabla_a \bm{\varphi} = (\partial_a - ie\bm{A}_a) \bm{\varphi},
\end{equation}
where $\partial_a$ is the coordinate derivative, $\bm{\varphi} \cong \varphi^\alpha$ is the gauge-charged field, $\alpha$ is the internal color index\footnote{If one expression is obtained from another by discarding indices, we prefer to connect them not by the equality $=$, but by the isomorphism sign $\cong$.

In our consideration, we do not use the choice of basis anywhere, so all indices can be understood as abstract. But, if it is more convenient for the reader, one can easily understand them in the usual sense as components, relative to the chosen basis.}, and $e$ is the coupling constant. Each component of the gauge field---vector potential $\bm{A}_a \cong A_{a\beta}{}^\alpha$ takes values in the corresponding Lie algebra. In the considered below case of $\mathrm{U}(n)$ symmetry this will be Lie algebra $\mathfrak{u}(n)$ of Hermitian $(n\times n)$-matrices\footnote{Or anti-Hermitian---it's just a matter of convention. It is known that in the complex case, Hermitian and anti-Hermitian matrices differ only by an imaginary unit $i$. In the physical literature it is customary to write $i$ in the formula \eqref{CovDerYM}, so that the potential $\bm{A}_a$ is Hermitian, while the mathematical tradition is opposite, so that $\bm{A}_a$ is anti-Hermitian.}, usually these matrices are decomposed over a basis of some normalized Lie algebra generators $\bm{A}_a = \bm{T}_i A_a^i$. In EG, a similar expression has the form
\begin{equation} \label{SpacetimeCovDer}
\nabla_a v^b = \partial_a v^b + \Gamma_{ac}{}^b v^c,
\end{equation}
where $\Gamma_{ac}{}^b$ are Christoffel symbols. Then the Yang-Mills field strength $\bm{F}_{ab} \cong F_{ab\alpha}{}^\beta$ and the Riemann tensor $R_{abc}{}^d$ are similarly constructed from the potentials $\bm{A}_a$ and the Christoffel symbols $\Gamma_{ac}{}^b$, respectively
\begin{align}
\bm{F}_{ab} &= \partial_a \bm{A}_b - \partial_b \bm{A}_a - ie\left[\bm{A}_a, \bm{A}_b\right], \label{BundleCurv} \\
R_{abc}{}^d & = \partial_a \Gamma_{bc}{}^d - \partial_b \Gamma_{ac}{}^d + \Gamma_{ah}{}^d \Gamma_{bc}{}^h - \Gamma_{bh}{}^d \Gamma_{ac}{}^h. \label{SpactimeCurv}
\end{align}

However, immediately after this similarity comes a fundamental difference: in YM, the dynamic field carrying physical degrees of freedom is the connection (i.e. the vector potential $\bm{A}_a$), and in standard EG the dynamic field is not the affine connection (i.e. Christoffel symbols $\Gamma_{ac}{}^b$), but the metric $g_{ab}$. The connection is the usual Levi-Civita connection of Riemannian geometry, fixed by conditions of torsionlessness and covariant constancy of the metric
\begin{equation} \label{CovConstMetric}
\nabla_a g_{bc} = 0.
\end{equation}

On a practical level, this is expressed by the fact that Yang-Mills equations are obtained by varying the Yang-Mills action
\begin{equation}
S_\mathrm{YM} = \frac{1}{4} \int d^dx\, \sqrt{g} \tr\left(\bm{F}_{ab}\bm{F}^{ab}\right)
\end{equation}
with respect to the potential $\bm{A}_a$, while the Einstein equations are obtained by varying the Hilbert-Einstein action
\begin{equation} \label{HEaction}
S_\mathrm{HE} = \frac{M_P^2}{2} \int d^dx\, \sqrt{g}\, R
\end{equation}
with respect to the metric $g_{ab}$\footnote{Throughout, we consider Euclidean field theory, in which the metric is positive definite. The transition to the physical Lorentzian case is accomplished by Wick rotation. We adopt the following conventions: $g = \det g_{ab}$, $R_{ab} = R_{acb}{}^c$, $R = g^{ab}R_{ab}$.}.

It seems quite natural that in order to bring these theories closer to each other, it is necessary to consider \emph{the connection and the metric as independent variables}. It is well-known that for the Hilbert-Einstein action \eqref{HEaction} the so-called ``Palatini approach'' does not lead to anything new, since varying $S_\mathrm{HE}$ with respect to $\Gamma_{ac}{}^b$ again leads to the covariant constancy condition \eqref{CovConstMetric} (\cite{MThWh}, see also \cite{PalatiniHistory, Palatini}). However, for other, more complicated gravitational actions (or even simply in the presence of matter fields) this is no longer the case: relaxing the condition \eqref{CovConstMetric} takes us beyond the standard Riemannian geometry, and we get some new theory, which is called \emph{metric-affine gravity (MAG)} \cite{Hehl1976, Buchbinder1992, Hehl1995, PBO2017, Baldazzi2022}. Note that since the geometry in this case is determined by a larger number of structures, there are more possibilities for constructing an action of such a theory, different from the standard action $S_\mathrm{HE}$. And, although MAG has been studied for decades, it can by no means be said that these models are fully understood.

However, our attention is now drawn to a completely different circumstance, which has always remained in shadows before. It is surprising that, as far as we know, no one has previously attempted to construct and study an analogous (and even much simpler)  \emph{``metric-affine-like'' generalization of the Yang-Mills theory (mal-YM)}. In this letter we show how such generalization can be constructed and point out some interesting properties of this theory.

\section{The key observation}

To construct the aforementioned generalization of YM, we must first ask the following simple question: if in MAG the partner of the affine connection $\Gamma_{ac}{}^b$ is the spacetime metric $g_{ab}$, then what structure must be the partner of the potential $\bm{A}_a$ in mal-YM? Perhaps the whole point is simply that for YM such a partner does not exist, and therefore the generalization we seek is impossible?

Turns out, this is not the case. As is well explained, for example, in the book by Penrose and Rindler (\cite{PenroseRindler1}, pp. 343-344), any specific symmetry in YM is equivalent to the presence of a certain additional structure in its internal color space (i.e. in the fibers of the corresponding vector bundle over spacetime). Unfortunately, the notation that is convenient for standard YM, and therefore has become widespread, leaves these structures hidden. This is probably the reason why a very simple generalization we suggest has not been proposed earlier. To build it, we need to bring the structure in fibers to the foreground, and for this we have to use a rather unusual Penrose notation \cite{PenroseRindler1}.

Since now we want, first of all, to show our general idea, for the sake of simplicity and clarity we will consider here and everywhere below one specific case---a theory with $\mathrm{U}(n)$ symmetry. Generalizations to other symmetry groups are straightforward. Also, for simplicity, we will assume that spacetime is flat everywhere below.

So, what additional structure in fibers corresponds to $\mathrm{U}(n)$ symmetry? In this case, the $n$-dimensional internal color space $V$ (a typical fiber of the bundle) will be complex. Note that the operation of complex conjugation cannot map the color space $V$ into itself, since otherwise it would contain real and imaginary vectors, but all vectors in the color space (on which the fundamental representation of the corresponding group acts) are equivalent. Therefore, complex conjugation will antilinearly map the color space into \emph{another space} $\bar V$, complex conjugate to it. Such complex conjugate spaces $V$ and $\bar V$ always appear in pairs. Following Penrose we will denote indices associated with $V$ and $\bar V$ by unprimed and primed Greek indices respectively\footnote{The undotted and dotted indices, standardly used to denote Weyl 2-spinors, have the same nature.}. Since unprimed and primed indices correspond to different spaces $V$ and $\bar V$, they cannot be contracted with each other. However, complex conjugation turns primed indices into unprimed ones and vice versa:
\begin{equation}
\overline{H_{\alpha\ldots\beta'\ldots}^{\gamma\ldots\delta'\ldots}} = \bar H_{\alpha'\ldots\beta\ldots}^{\gamma'\ldots\delta\ldots}
\end{equation}

Now, suppose we are given a $\mathrm{U}(n)$-charged scalar $\bm{\varphi} \cong \varphi^\alpha$. How can we construct a covariant real Lagrangian density for it? Obviously, this requires not only the complex conjugate scalar $\bar\varphi^{\alpha'}$, but also a form in the fibers $g_{\alpha\beta'}$, which is Hermitian and non-degenerate
\begin{equation} \label{gDef}
\bar g_{\alpha\beta'} = g_{\alpha\beta'}, \qquad g_{\alpha\gamma'}g^{\beta\gamma'} = \delta_\alpha^\beta.
\end{equation}
Then, using it, the desired Lagrangian density can be written as
\begin{equation} \label{LagPhi}
\calL_\varphi = \frac{1}{2} g_{\alpha\beta'} \nabla_a\varphi^\alpha \nabla^a \bar\varphi^{\beta'} + P\left(|\varphi|^2\right),
\end{equation}
where $|\varphi|^2 = g_{\alpha\beta'}\varphi^\alpha \bar\varphi^{\beta'}$, and $P\left(|\varphi|^2\right)$ is some potential bounded from below, which describes self-interaction of the field $\bm{\varphi}$.

We claim that this Hermitian form in fibers $g_{\alpha\beta'}$ is the very analogue of the spacetime metric $g_{ab}$, which will play the role of potential partner in mal-YM. Indeed, $g_{\alpha\beta'}$ is quite similar in its properties to $g_{ab}$: Hermiticity is an analogue of the symmetry of the metric, and non-degeneracy allows raising and lowering color indices in the usual way. In this case, when changing the position, the unprimed index becomes primed and vice versa, for example, $\bar\varphi_\alpha = g_{\alpha\alpha'}\bar\varphi^{\alpha'}$. Thus, the Hermitian form $g_{\alpha\beta'}$ establishes a (non-canonical) isomorphism $\bar V \cong V^*$ between the complex conjugate and the dual spaces.

Moreover, the analogy between YM and EG can be carried much further: in fact, the standard YM also has an implicit assumption that the Hermitian form in the fibers is covariantly constant
\begin{equation} \label{CovConstYM}
\nabla_a g_{\alpha\beta'} = 0,
\end{equation}
which is completely analogous to the condition of covariant constancy of the spacetime metric \eqref{CovConstMetric} in EG. Accordingly, \emph{the ``metric-affine-like'' generalization of YM can be obtained by dropping this condition}.

In the remainder of this letter we briefly, without any detailed explanations, list the main consequences of this generalization. All the statements made below will be substantiated in detail in a separate article \cite{Wachowski24b}.

\section{Hermitian and anti-Hermitian parts and gauge transformations}

Let us agree to call ``matrices'' linear maps of the internal color space $V\to V$, i.e. tensors with two unprimed indices---one upper and one lower $\bm{M} \cong M_\alpha^\beta$, so that the product of matrices is simply a contraction over one of the indices $\bm{M}\bm{N} \cong M^\alpha_\gamma N^\gamma_\beta$. Let us note the following: the complex conjugate object $\bar M_{\alpha'}^{\beta'}$ is no longer a ``matrix'' in this sense---it is a map of the complex conjugate space $\bar V\to \bar V$, it has primed indices, and therefore cannot be multiplied by ``matrices''. However, we can define the operation of \emph{Hermitian conjugation}, which maps matrices to matrices, but contains not only complex conjugation, but also ``transposition,'' i.e. contractions with $g_{\alpha\alpha'}$ and $g^{\beta\beta'}$:
\begin{equation} \label{HermConjDef}
\bm{M}^\dag \cong \bar M_\alpha^\beta = g_{\alpha\alpha'} \bar M_{\beta'}^{\alpha'} g^{\beta\beta'}.
\end{equation}
Although it is not explicitly stated in standard algebra courses, Hermitian conjugation always requires the specification of some Hermitian form $g_{\alpha\beta'}$.

Let us be given two different connections $\nabla_a$ and $\tilde\nabla_a$. We define the \emph{total} curvature in the bundle $\calF_{ab}[\nabla] \cong \scrF_{ab}{}_\alpha^\beta$ and the \emph{total} connection potential $\calA_a[\tilde\nabla-\nabla] \cong \scrA_a{}_\alpha^\beta$ by the relations\footnote{$\calA_a$ defined in this way is not the usual vector potential, but a difference (or variation) of potentials. Accordingly, $\bm{A}_a$ below differs from the potential in formulas~\eqref{CovDerYM} and~\eqref{BundleCurv}.}
\begin{align}
&[\nabla_a, \nabla_b] \bm{\varphi} = \calF_{ab} \bm{\varphi}, \\
&(\tilde\nabla_a - \nabla_a) \bm{\varphi} = \calA_a \bm{\varphi},
\end{align}
where $\bm{\varphi} \cong \varphi^\alpha$. Then the action of $[\nabla_a, \nabla_b]$ and $(\tilde\nabla_a - \nabla_a)$ on other gauge-charged (i.e. carrying color indices) fields is obtained simply from the Leibniz rule.

Using Hermitian conjugation \eqref{HermConjDef}, we can split $\calA_a$ and $\calF_{ab}$ into Hermitian and anti-Hermitian parts
\begin{equation}
\calA_a = \tilde e\bm{B}_a -  ie \bm{A}_a, \qquad \calF_{ab} = \tilde e\bm{G}_{ab} -  ie \bm{F}_{ab},
\end{equation}
where $e$ and $\tilde e$ are two positive coupling constants, and
\begin{align}
\bm{B}_a &= \frac{1}{2\tilde e}(\calA_a + \calA_a^\dag), & \bm{A}_a &= \frac{i}{2e}(\calA_a - \calA_a^\dag), \label{TotalPotential} \\
\bm{G}_{ab} &= \frac{1}{2\tilde e}(\calF_{ab} + \calF_{ab}^\dag), & \bm{F}_{ab} &= \frac{i}{2e}(\calF_{ab} - \calF_{ab}^\dag). \label{TotalCurvature}
\end{align}

As we mentioned at the very beginning, in YM only anti-Hermitian components $\bm{F}_{ab}$ and $\bm{A}_a$ are present. Since this statement always appears at the very beginning and is not substantiated in any way, it may seem that this is some kind of axiom. But why should we adhere to it? In fact, this is not an axiom at all, but a very simple consequence of a deeper condition of covariant constancy of the structure in fibers \eqref{CovConstYM}. However, if we abandon this condition, it will no longer be the case.

Let us introduce the following quantity, which is a measure of violation of the condition~\eqref{CovConstYM} and which we will therefore call \emph{the YM-deviation vector}:
\begin{equation} \label{YMdeviationDef}
\bm{N}_a \cong N_{a\alpha}{}^\beta = -\frac{1}{2\tilde e} g^{\beta\beta'} \nabla_a g_{\alpha\beta'}.
\end{equation}
It can be shown that the Hermitian part of the total curvature $\bm{G}_{ab}$ is completely determined by the YM-deviation vector $\bm{N}_a$:
\begin{equation} \label{NtoG}
\bm{G}_{ab} = \nabla_a \bm{N}_b - \nabla_b \bm{N}_a - 2\tilde e [\bm{N}_a, \bm{N}_b].
\end{equation}
If the condition \eqref{CovConstYM} is satisfied, then the YM-deviation vector vanishes $\bm{N}_a = 0$, so $\bm{G}_{ab} = 0$, and the total curvature in the bundle is reduced to the usual field strength tensor $\calF_{ab} = -ie \bm{F}_{ab}$. Thus, we restore the standard YM. Another way to formulate this is as follows: \emph{if a connection respects the Hermitian form in fibers, then its total curvature will be anti-Hermitian}.

Note that this situation is an (even somewhat simpler) analogy to what happens in MAG. There, if we abandon the condition of covariant constancy of the metric \eqref{CovConstMetric}, the spacetime curvature ceases to be anti-symmetric with respect to the second pair of indices, but its symmetric part $R_{ab(cd)}$ is completely determined by the non-metricity $\nabla_a g_{cd}$ and vanishes together with it, which returns us to standard EG.

Now, let $\bm{u} \cong u_\alpha^\beta$ and $\bm{U} \cong U_\alpha^\beta$ be two arbitrary mutually inverse complex matrices:
\begin{equation}
\bm{u}\bm{U} = \bm{U}\bm{u} = \mathbf{1}.
\end{equation}
Consider general linear transformations of the color space together with the corresponding change of connection
\begin{align}
&\calA_a[\tilde\nabla-\nabla] = \bm{U} \nabla_a\bm{u}, \label{PureGaugePotencalEq} \\
&H_{\alpha\ldots}^{\beta\ldots} \mapsto \tilde H_{\alpha\ldots}^{\beta\ldots} = U_\gamma^\beta\cdots\; H_{\delta\ldots}^{\gamma\ldots}\; u_\alpha^\delta \cdots. \label{InvColorTransform}
\end{align}
Since such transformations obviously preserve any index contractions, they will represent $\mathrm{GL}(n,\mathbb{C})$ gauge symmetry of any geometrically defined theory.

However, introducing the Hermitian form $g_{ \alpha\beta'}$ breaks this symmetry to $\mathrm{U}(n)$. Indeed, the Hermitian form under transformations \eqref{InvColorTransform} will change as
\begin{equation}
\tilde g_{\alpha\beta'} = \omega_\alpha^\beta g_{\beta\beta'}, \quad\text{where}\quad \omega_\alpha^\beta \cong \bm{\omega} = \bm{u}^\dag \bm{u}.
\end{equation}
So the requirement that $g_{\alpha\beta'}$ must be invariant singles out a narrower class of \emph{unitary transformations} for which $ \bm{U} = \bm{u}^\dag$. It is convenient to parameterize the matrices $\bm{u}$ and $\bm{\omega}$ as follows
\begin{equation} \label{ExpParametrization}
\bm{u} = \exp(\bm{\epsilon}), \qquad \bm{\omega} = \exp(\bm{h}),
\end{equation}
and decompose the gauge transformation parameter into Hermitian and anti-Hermitian parts $\bm{\epsilon} = \tilde e\bm{\beta} - ie \bm{\alpha}$. Then, for infinitesimal transformations, we have $\bm{h} = 2\tilde e\bm{\beta}$. Thus, $\bm{\alpha}$ corresponds to usual unitary transformations preserving the Hermitian form $g_{\alpha\beta'}$, and $\bm{\beta} $ corresponds to non-unitary transformations that change it. Therefore, $g_{\alpha\beta'}$ is a Higgs-like field, and its variation $\bm{h}$ is a Goldstone boson, or compensator, or St\"{u}ckelberg field~\cite{Ruegg2004}.

The broken $\mathrm{GL}(n, \mathbb{C})$ gauge invariance also manifests itself in the Noether identities. We define anti-Hermitian $\bm{J}_a$ and Hermitian $\bm{\Lambda}_a$ gauge currents as variations of the action of the theory $S[\nabla, g]$ with respect to the anti-Hermitian $\bm{A}_a$ and Hermitian $\bm{B}_a$ potentials, respectively, and we also define the Hermitian matrix $\bm{E} = \bm{E}^\dag$, which is the source of the field $\bm{h}$
\begin{equation}
\bm{J}_a = \frac{\delta S}{\delta\bm{A}^a}, \quad \bm{\Lambda}_a = -\frac{\delta S}{\delta\bm{B}^a}, \quad \bm{E} = -2\frac{\delta S}{\delta\bm{h}}.
\end{equation}
Then in pure mal-YM without matter fields, the Noether identities following from the symmetry under unitary $\bm{\alpha}$ and non-unitary $\bm{\beta}$ gauge transformations, respectively, will be of the form:
\begin{align}
&\nabla_a\bm{J}^a - \tilde e[\bm{N}_a, \bm{J}^a] + ie [\bm{N}_a, \bm{\Lambda}^a] = 0, \\
&\nabla_a\bm{\Lambda}^a - \tilde e[\bm{N}_a, \bm{\Lambda}^a] - i\frac{\tilde e^2}{e} [\bm{N}_a, \bm{J}^a] = \tilde e\bm{E}.
\end{align}
The first of these relations is a modification of the conservation law of the Yang-Mills current, which curiously ceases to be preserved in mal-YM. And the second relation means that the equation obtained by varying the action with respect to $\bm{h}$ is not independent, but only a differential consequence of other EoMs.

\section{Action and equations of motion}

Summarizing the above, now in our theory in addition to the usual Yang-Mills field $\bm{F}_{ab}$, $\bm{A}_a$ there appears new gauge fields $\bm{G}_{ab}$ and $\bm{N}_a$, $\bm{B}_a$ and $\bm{h}$, which can be considered a non-Abelian generalization of the St\"{u}ckelberg theory \cite{Ruegg2004}. The dynamics of these fields and their interactions with each other are determined by the form of the action.

We will consider the theory with the Lagrangian
\begin{equation} \label{totalLagrangian}
\calL_\text{mal-YM} = \frac{1}{4}\tr(\bm{F}_{ab} \bm{F}^{ab} + \bm{G}_{ab}\bm{G}^{ab} + 2M^2 \bm{N}_a \bm{N}^a).
\end{equation}
Here the coefficient before the first two terms is chosen so that in the quadratic in perturbations part of the action terms $\nabla_a\bm{A}_b\nabla^a\bm{A}^b$ and $\nabla_a\bm{B}_b\nabla^a\bm{B}^b$ have the correct coefficient $1/2$. The introduction of the third mass term is motivated by the natural desire to be able to restore the condition \eqref{CovConstYM} and, consequently, the standard YM  in limit $M\to\infty$. Of course, we could also introduce other terms, for example, $\tr(\bm{G}_{ab}\bm{F}^{ab})$, $\tr([\bm{N}_a, \bm{N}_b]\bm{F}^{ab})$, etc., but we do not do this solely for the sake of simplicity, in order not to clutter the presentation. It is also convenient to redefine the variation of the Hermitian form~\eqref{ExpParametrization} $\bm{h} \mapsto 2\tilde e\bm{h}/M$.

The Lagrangian \eqref{totalLagrangian} leads to the following classical equations on the background fields:
\begin{align}
&\nabla^b \bm{F}_{ab} - \tilde e [\bm{N}^b, \bm{F}_{ab}] - ie [\bm{N}^b, \bm{G}_{ab}] = -\bm{J}_a^\mathrm{ext}, \label{EqAfinal} \\
&\nabla^b \bm{G}_{ab} - \tilde e [\bm{N}^b, \bm{G}_{ab}] + i\frac{\tilde e^2}{e} [\bm{N}^b, \bm{F}_{ab}] + M^2\bm{N}_a = \bm{\Lambda}_a^\mathrm{ext}, \label{EqBfinal}
\end{align}
where on the right-hand side we have added external anti-Hermitian $\bm{J}_a^\mathrm{ext}$ and Hermitian $\bm{\Lambda}_a^\mathrm{ext}$ gauge currents that can be generated by the matter fields if they are present in the theory. For example, for a gauge-charged scalar field with Lagrangian \eqref{LagPhi} they will have the form:
\begin{align}
&\bm{J}_a \cong J_{a\alpha}{}^\beta = \frac{ie}{2} g_{\alpha\beta'} \left(\bar\varphi^{\beta'} \nabla_a \varphi^\beta - \varphi^\beta \nabla_a \bar\varphi^{\beta'}\right), \label{CurrentPhiA} \\
&\bm{\Lambda}_a \cong \Lambda_{a\alpha}{}^\beta = -\frac{\tilde e}{2} g_{\alpha\beta'} \nabla_a (\bar\varphi^{\beta'} \varphi^\beta). \label{CurrentPhiB}
\end{align}

Finally, we also present equations for small perturbations $\bm{A}_a$, $\bm{B}_a$, and $\bm{h}$, propagating on a trivial background $\bm{N}_a = 0$, $\bm{F}_{ab} = \bm{G}_{ab} = 0$. For the usual YM potential $\bm{A}_a$, we obtain the standard equation for a massless vector boson
\begin{equation} \label{LinEqA}
(\delta_a^b \Box + \partial^b\partial_a) \bm{A}_b = 0,
\end{equation}
where $\Box = -\partial^a\partial_a$ is the flat Laplacian. And the equations for the fields $\bm{B}_a$ and $\bm{h}$ are similar to those that arise in St\"{u}ckelberg theory \cite{Ruegg2004}:
\begin{align}
&(\delta_a^b \Box + \partial^b\partial_a)\bm{B}_b + M(M\bm{B}_a - \partial_a\bm{h}) = 0, \label{LinEqB} \\
&\Box\bm{h} + M\partial_a\bm{B}^a = 0. \label{LinEqH}
\end{align}
Clearly, the second of these equations is a differential consequence of the first, and both are invariant under non-unitary gauge transformations
\begin{equation}
\delta\bm{B}_a = \partial_a\bm{\beta}, \qquad \delta\bm{h} = M\bm{\beta}.
\end{equation}
If we fix the ``unitary'' gauge $\bm{h} = 0$, then the field $\bm{B}_a$ obeys the Proca equation and is a massive vector boson with mass $M$.

It can be noted that if we look for solutions of the equations \eqref{EqAfinal}-\eqref{EqBfinal} without external sources, for which $\bm{N}_a = 0$ (and hence also $\bm{G}_{ab} = 0$), then the remaining field $\bm{F}_{ab}$ will satisfy the usual Yang-Mills equation $\nabla^b\bm{F}_{ab} = 0$. Thus, \emph{any classical solution of pure YM is at the same time a solution of pure mal-YM}. However, it apparently follows from the equations \eqref{LinEqB}-\eqref{LinEqH} that, at least in the vicinity of the trivial background, mal-YM has solutions that are not borrowed from YM.

It also follows from the equation \eqref{EqBfinal} that when the mass of the second field tends to infinity, $M\to\infty$, we obtain $\bm{N}_a = 0$, i.e. we restore the covariant constancy condition \eqref{CovConstYM} and return to the standard YM.

\section{Potential significance of mal-YM}

To summarize, we have constructed a very natural generalization of YM, in which, in addition to the usual massless Yang-Mills gauge fields $\bm{A}_a$ and $\bm{F}_{ab}$, there appear new non-Abelian St\"{u}ckelberg fields $\bm{B}_a$ and $\bm{h}$, $\bm{G}_{ab}$ and $\bm{N}_a$, which can be made massive. Letting their mass tend to infinity $M\to\infty$ restores the standard YM.

Here we have considered this theory exclusively at the classical level. The question is whether it is also good at the quantum level, in particular, renormalizable. Indeed, as we discuss in more detail in \cite{Wachowski24b}, this theory is non-polynomial in the field $\bm{h}$, i.e., it contains Feynman diagrams with vertices of arbitrarily high order and, accordingly, with couplings of arbitrarily large negative dimension. Using the broken $\mathrm{GL}(n, \mathbb{C})$ gauge invariance, we can pass to the $\bm{h} = 0$ gauge, completely eliminating non-polynomial interactions. But then $\bm{B}_a$ becomes a Proca field whose propagator does not decrease in the UV region. We can fix the propagators by switching to the Feynman-'t Hooft gauge, but this introduces a nonzero field $\bm{h}$ and the associated non-polynomiality.

Therefore, the question of renormalizability and other quantum properties of mal-YM is quite subtle and requires a separate careful study. The potential theoretical significance of such a study could be that it would simultaneously allow us to better understand both the ordinary YM and its underlying covariant constancy condition \eqref{CovConstYM}, as well as MAG and its difficulties. Therefore, even if it turns out that mal-YM for some reason cannot be considered physical, this would already be an interesting result.

However, if mal-YM turns out to be a physically admissible theory, then such a new generalization with additional parameters has great phenomenological interest. Indeed, the fact that we do not observe the fields $\bm{B}_a$ and $\bm{G}_{ab}$ can always be explained by mass $M$ is too large. However, this immediately leads us to the question of what is the lower bound to this mass compatible with the data of collider experiments and observational cosmology? Further, can the new theory help explain some known phenomena? And can it be that some of the periodically raising (and usually dissolving) deviations from the Standard Model are better described by the new theory at some values of its parameters $e$, $\tilde e$ and $M$?

\section*{Acknowledgments}

The author expresses deep gratitude to his colleagues A.\,O.~Barvinsky, A.\,E.~Kalugin, N.\,M.~Kolganov, A.\,V.~Kurov and D.\,V.~Nesterov for interesting and stimulating discussions. The research was supported by the Russian Science Foundation grant No. 23-12-00051, \url{https://rscf.ru/en/project/23-12-00051/}.

\bibliographystyle{apsrev4-2}
\bibliography{Wachowski2605}

\end{document}